\documentclass[lettersize,journal]{IEEEtran}
\usepackage{amsmath,amsfonts}
\usepackage{algorithmic}
\usepackage{algorithm}
\usepackage{array}
\usepackage[caption=false,font=normalsize,labelfont=sf,textfont=sf]{subfig}
\usepackage{textcomp}
\usepackage{stfloats}
\usepackage{url}
\usepackage{verbatim}
\usepackage{graphicx}
\usepackage{cite}
\usepackage[all=normal, floats, bibnotes, wordspacing, charwidths, indent, lists]{savetrees}

\usepackage{bbm}
\usepackage{booktabs}
\usepackage{multirow}
\usepackage{xspace}
\usepackage{mathtools}
\usepackage{amsthm}
\usepackage{hyperref}
\usepackage{color}

\graphicspath{{./figs/}}

\theoremstyle{plain}
\newtheorem{theorem}{Theorem}
\newtheorem{assumption}[theorem]{Assumption}

\usepackage[table]{xcolor}
\definecolor{light-gray}{gray}{0.9}

\makeatletter

\newcommand{\Rmnum}[1]{\expandafter\@slowromancap\romannumeral #1@}
\makeatother

\newcommand{\figref}[1]{\figurename~\ref{#1}}
\newcommand{\tabref}[1]{\tablename~\ref{#1}}
\newcommand{\algref}[1]{Algorithm~\ref{#1}}
\newcommand{\secref}[1]{Section~\ref{#1}}

\newcommand{\bheading}[1]{{\vspace{0.2\baselineskip}\noindent{\textbf{#1}}}}
\newcommand{\lheading}[1]{{\vspace{0.2\baselineskip}\noindent{\textit{#1}}}}

\newcommand{\eg}{\textit{e.g.}\xspace}
\newcommand{\ie}{\textit{i.e.}\xspace}
\newcommand{\etal}{\textit{et~al.}\xspace}

\begin{document}

\title{Exploring the Adversarial Frontier: Quantifying Robustness via Adversarial Hypervolume}

\author{
    Ping Guo, Cheng Gong, Xi Lin, Zhiyuan Yang, Qingfu Zhang\\
    Department of Computer Science, City University of Hong Kong, Hong Kong\\
}



\maketitle

\begin{abstract}
    The escalating threat of adversarial attacks on deep learning models, particularly in security-critical fields, has highlighted the need for robust deep learning systems. Conventional evaluation methods of their robustness have relied on adversarial accuracy, which measures the model performance under a specific perturbation intensity. However, this singular metric does not fully encapsulate the overall resilience of a model against varying degrees of perturbation. To address this issue, we propose a new metric termed as the adversarial hypervolume for assessing the robustness of deep learning models comprehensively over a range of perturbation intensities from a multi-objective optimization standpoint. This metric allows for an in-depth comparison of defense mechanisms and recognizes the trivial improvements in robustness brought by less potent defensive strategies. We adopt a novel training algorithm to enhance adversarial robustness uniformly across various perturbation intensities, instead of only optimizing adversarial accuracy. Our experiments validate the effectiveness of the adversarial hypervolume metric in robustness evaluation, demonstrating its ability to reveal subtle differences in robustness that adversarial accuracy overlooks.
\end{abstract}


\section{Introduction}\label{sec:intro}
The increasing emphasis on evaluating the inherent robustness of deep learning (DL) models and developing robust DL systems constitutes a pivotal area of research within the machine learning (ML) community~\cite{madry:2018:towards,ren:2018:learning,salman:2019:provably,dong:2020:adversarial,Sadeghi:2020:taxonomy}.
This trend is primarily driven by growing security concerns with \emph{adversarial attacks} on deep neural network-based (DNN) image classification systems. These attacks introduce imperceptible perturbations to input images, which are typically indistinguishable from the human eye, leading to incorrect classifications by the DNN model~\cite{szegedy:2014:intriguing,goodfellow:2015:explaining,kurakin:2017:adversarial}.
Such vulnerabilities are particularly alarming in contexts critical to security, such as face recognition~\cite{dong:2019:efficient} and autonomous driving~\cite{cao:2019:adversarial}, where the consequences of misclassification could be severe.
Nonetheless, efforts to construct adversarially robust models face substantial challenges, particularly the absence of a comprehensive evaluation metric or framework for evaluation of models' robustness.

Conventionally, \emph{adversarial accuracy} has been used as the principal metric for robustness evaluation, measuring a model's resilience to adversarial examples generated through a range of attack strategies~\cite{goodfellow:2015:explaining,carlini:2017:towards,athalye:2018:obfuscated}.
The emergence of increasingly sophisticated attacks, such as the Fast Gradient Sign Method (FGSM)~\cite{goodfellow:2015:explaining}, DeepFool~\cite{dezfooli:2016:deepfool}, Projected Gradient Descent (PGD)~\cite{madry:2018:towards}, and Auto Projected Gradient Descent (APGD)~\cite{croce:2020:reliable}, has led to an expansion of benchmark of adversarial accuracies. Despite this, the field lacks a unified robustness metric due to the differing adversarial accuracy results across various attacks.
Notably, the integration of APGD~\cite{croce:2020:reliable} and Square Attack~\cite{andriushchenko:2020:square} within AutoAttack~\cite{croce:2020:reliable} has established a widely recognized benchmark for adversarial accuracy assessment.
The development of standard toolkits like RobustBench~\cite{croce:2021:robustbench} has also substantially advanced the uniform comparison of adversarial defense capabilities across different defensive models, positioning the adversarial accuracy of the models under AutoAttack as the gold standard.

However, recent research has challenged the adequacy of adversarial accuracy as the robustness metric for DL models, demonstrating that it fails to capture the full scope of a model's resilience against attacks~\cite{carlini:2017:towards,robey:2022:probabilistically,oliver:2023:howmany}.
Robey~\etal~\cite{robey:2022:probabilistically} have proposed to use the probability of misclassification as a more comprehensive measure of robustness, calculated through random sampling, although they acknowledge the computational challenges posed by the complexity of high-dimensional spaces.
Oliver~\etal~\cite{oliver:2023:howmany} have explored non-adversarial spaces using geometric methods and assessed the extent of these regions within the sample space as an indicator of robustness.
Despite these contributions, one can argue that neither misclassification probability nor adversarial region size sufficiently represent a model's robustness, as they fail to consider the variation in a model's response to varying levels of perturbation intensities.

\begin{figure*}[t]
  \centering
  \includegraphics[width=0.88\textwidth]{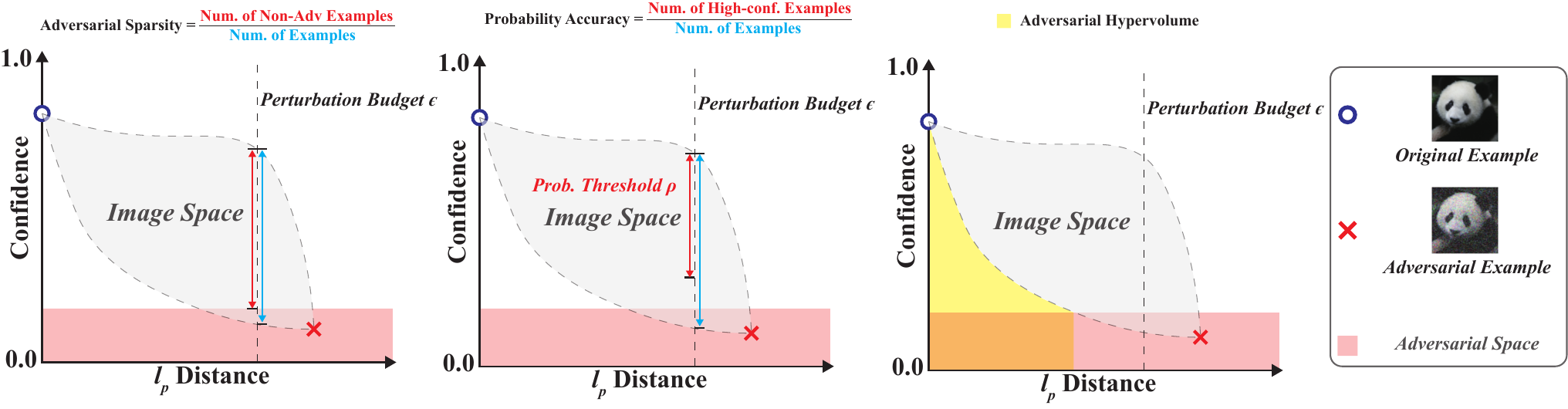}
  \vspace{-1\baselineskip}
  \caption{\textbf{Comparison between adversarial hypervolume and established robustness metrics.} \emph{Adversarial sparsity} measures the proportion of non-adversarial to total examples at perturbation level $\epsilon$. \emph{Probability accuracy} denotes the ratio of high-confidence predictions to total examples at the perturbation level $\epsilon$. \emph{Adversarial hypervolume} represents the averaged variations in confidence scores over a range of perturbation intensities.\label{fig:compare}}
  \vspace{-1.25\baselineskip}
\end{figure*}

To overcome the shortcomings of existing robustness metrics, which focus solely on model robustness at a fixed perturbation level, we introduce a new metric called \emph{adversarial hypervolume}.
This metric evaluates models' robustness by examining the normalized confidence scores of worst-case adversarial examples across various levels of perturbation intensities.
It facilitates meaningful comparisons between defensive models with almost same adversarial accuracy and evaluation of the incremental robustness improvements provided by weaker defensive measures like Feature Squeezing and JPEG compression~\cite{xu:2018:feature}.
A higher adversarial hypervolume metric value indicates higher model robustness throughout various perturbation intensities, thereby offering a more comprehensive assessment of a model's resilience against adversarial attacks.
We illustrate a conceptual comparison between our proposed metric and two existing metrics in \figref{fig:compare}.

Our proposed metric, adversarial hypervolume, is based on the concept of hypervolume~\cite{bader:2011:hype} from the multi-objective optimization literature~\cite{deb:2016:multi,deb:2002:nsga2,zhang:2007:moead}.
We begin by constructing a multi-objective optimization problem that mirrors the complexity of actual attack scenarios, with the dual aims of minimizing the perturbation's magnitude and simultaneously reducing the model's confidence in the original classification.
In contrast, much of existing research approaches the problem from the perspective of single-objective optimization with a distance constraint~\cite{goodfellow:2015:explaining, madry:2018:towards} or employs a weighted sum of multiple objectives~\cite{carlini:2017:towards}.
We subsequently delineate a set of adversarial examples that illustrate the model's unique vulnerability landscape when subjected to an attack. Each example signifies a nadir of score corresponding to a particular magnitude of perturbation, thereby delineating the \emph{adversarial frontier}.
Finally, we quantify a model's robustness by calculating the adversarial hypervolume, which measures the size of the region enclosed by the adversarial frontier.

In summary, our contributions are as follows:
\begin{itemize}
  \item We introduce a new metric, the adversarial hypervolume, which quantifies the robustness of DL models across a range of perturbation intensities, thereby offering a comprehensive evaluation of a model's resilience to adversarial attacks.
  \item Utilizing this metric, we devise a straightforward method for its computation and implement an efficient training algorithm to improve the robustness of DL models.
  \item Our empirical findings reveal that the adversarial hypervolume effectively measures the enhancement of robustness from weaker defensive mechanisms (\ie, input transformations) and delivers important supplementary insights, such as average confidence variation, which augment the understanding provided by adversarial accuracy.
  \item We conduct extensive experiments to assess the robustness of cutting-edge state-of-the-art (SOTA) defensive models, establishing a benchmark for subsequent research.
\end{itemize}

\section{Related Work}
\subsection{Adversarial Attacks and Defenses}
\bheading{Adversarial Attacks.}
Following the identification of adversarial vulnerabilities in DL models~\cite{szegedy:2014:intriguing}, there has been a significant trend in research dedicated to the creation of adversarial attack algorithms. These studies explored to generate adversarial examples within different threat models~\cite{akhtar:2018:threat,machado:2023:adversarial,li:2023:trustworthy}.
Specifically, with different levels of information of the models, researchers have developed \emph{white-box} attacks and \emph{black-box} attacks respectively.

\lheading{White-box Attacks.}
White-box attacks formulates an optimization problem aimed at creating adversarial examples by leveraging full model knowledge, \eg, gradient information. This approach has given rise to various attack methodologies, notably the PGD attack~\cite{madry:2018:towards}, the Carlini-Wagner (CW) attack~\cite{carlini:2017:towards}, the FGSM attack~\cite{goodfellow:2015:explaining}, the Basic Iterative Method (BIM) attack~\cite{kurakin:2017:adversarial}, and the DeepFool attack~\cite{dezfooli:2016:deepfool}.

\lheading{Black-box Attacks.}
Black-box attacks operate under a threat model where only the model's input and output are known~\cite{bhambri:2019:survey,li:2023:saes,li:2020:qeba,fu:2022:autoda,guo:2024:lautoda,zanddizari:2022:gener}.
These attacks circumvent the limitations of white-box attacks in realistic scenarios where the model's internal details are inaccessible.
Notably, the Square attack~\cite{andriushchenko:2020:square}, despite relying on less information than white-box attacks, proves to be more effective than some white-box methods.
It has been incorporated into the AutoAttack framework~\cite{croce:2020:reliable} for robustness evaluation.

\lheading{Attacks for Evaluating Robustness.}
Employing strong attacks to measure adversarial accuracies is a standard practice in robustness evaluation of DL models~\cite{madry:2018:towards,croce:2020:reliable}.
Notably, the PGD attack~\cite{madry:2018:towards} and the AutoAttack framework~\cite{croce:2020:reliable} stand as the most widely adopted attacks for such evaluations. AutoAttack combines the APGD attack with the Square attack.
Moreover, the collection of the attack strategies is rapidly growing, now encompassing state-of-the-art approaches such as the Composite Adversarial Attack (CAA)~\cite{mao:2021:composite} and AutoAE~\cite{liu:2023:reliable}.

\bheading{Adversarial Defenses.}
Defensive strategies against adversarial attacks are designed to enhance a model's inherent robustness or alleviate the effects of adversarial attacks.
To improve the models' robustness, a range of adversarial training techniques have been introduced~\cite{madry:2018:towards,bai:2021:recent}, along with fine-tuning on dataset or models~\cite{ye:2024:shed,wang:2024:flora,liu:2023:twins}.
Furthermore, several tactics have been devised to detect or neutralize adversarial examples, including input transformation~\cite{guo:2018:countering,xu:2018:feature}, adversarial purification~\cite{guo:2024:puridefense,carlini:2023:certified}, and stateful defenses~\cite{ma:2019:nic,li:2022:blacklight}.
Given that adversarial purification and stateful defenses are tailored for real-world classification systems rather than a standalone model, our study concentrates primarily on the robustness improvement brought by adversarial training and input transformation techniques.

\lheading{Adversarial Training.}
Adversarial training is a widely adopted method to enhance the robustness of DL models by integrating adversarial examples into the training dataset. This integration facilitates the model in developing more robust decision boundaries.
Adversarial examples are often generated using the PGD attack~\cite{madry:2018:towards}, a substantial body of research supports its efficacy~\cite{zhang:2019:theoretically,wang:2023:better,gowal:2020:uncovering}.

\lheading{Input Transformation.}
Input transformation is a cost-effective defensive strategy that can be seamlessly incorporated into systems with adversarially trained models to enhance robustness~\cite{guo:2018:countering,xu:2018:feature}.
However, the efficacy of input transformation has come into question, as it has been proven to be vulnerable to adaptive attacks~\cite{athalye:2018:obfuscated}.
Moreover, using adversarial accuracy to measure the robustness improvement from input transformation techniques may lead to misleading results, since the adversarial accuracy of disparate techniques tends to be low and indistinguishable.
In contrast, our research introduces the adversarial hypervolume metric, which facilitates a comparative analysis of robustness improvements by providing more detailed insights into an averaged version of confidence value.

\subsection{Metrics for Evaluating Robustness}
To gain a deeper insight into the robustness of models, researchers have introduced a variety of new metrics that extend beyond merely adversarial accuracy. Such metrics include the minimum adversarial perturbation~\cite{carlini:2017:towards}, probabilistic robustness~\cite{robey:2022:probabilistically}, and adversarial sparsity~\cite{oliver:2023:howmany}.

\bheading{Minimal Perturbation.}
Quantifying the minimal magnitude of perturbation necessary to transform an input into an adversarial example is a pioneering attempt toward robustness evaluation~\cite{szegedy:2014:intriguing}. Carlini~\etal approached this issue by formulating an optimization problem that strikes a balance between perturbation magnitude and the confidence of the adversarial example, incorporating their linear combination~\cite{carlini:2017:towards}. However, this metric does not comprehensively reflect the full range of a model's robustness across various perturbation levels.

\bheading{Probabilistic Robustness.}
Robey~\etal proposed a novel metric for evaluating probabilistic robustness, thereby redirecting attention away from traditional \textit{worst-case} scenarios~\cite{robey:2022:probabilistically}.
This metric assesses the probability of maintaining accurate predictions when a model is subjected to adversarial perturbations.
However, one limitation is the considerable computational demand, typically requiring Monte Carlo simulations to estimate the distribution of adversarial examples, which becomes particularly challenging in high-dimensional settings.

\bheading{Adversarial Sparsity.}
Recent research by Oliver~\etal has presented the concept of adversarial sparsity, which probes the boundary of non-adversarial regions to determine their proportion within the entire sample space~\cite{oliver:2023:howmany}.
Although the method aligns with the principles of probabilistic robustness, Oliver~\etal suggest a geometric method to determine the adversarial sparsity, which yields improved convergence results.

\subsection{Multiobjective Optimization in Adversarial Robustness}
Within the field of adversarial robustness, multiobjective optimization serves two main purposes: the enhancement of robustness and the development of a multiobjective adversarial framework.
The former enhances model resilience against attacks by incorporating multiobjective optimization to refine the traditional norm-bounded problem.
In contrast, the latter introduces new problem frameworks that more closely align with our research aims.

\bheading{Multiobjecive Optimization for Robustness.}
Recent research efforts have advanced the application of multiobjective optimization in crafting adversarial examples.
These efforts have expanded the original single-objective paradigm to include additional aims, yielding more diverse and robust adversarial examples~\cite{williams:2023:blackbox,williams:2023:sparse}. Additionally, multiobjective optimization has been employed in training models, offering a defensive enhancement to adversarial robustness~\cite{deist:2023:multi}.

\bheading{Multiobjecive Adversarial Problem.}
The multiobjective adversarial problem introduces an innovative optimization framework to the domain of adversarial robustness, targeting the enhancement of model robustness across a range of perturbation intensities~\cite{Suzuki:2019:adversarial,baia:2021:effective,liu:2024:effective}.
While Suzuki~\etal's~\cite{Suzuki:2019:adversarial} analysis aligns with our investigation, it restricts its scope to preliminary results with the VGG16 model.
In contrast, Baia~\etal~\cite{baia:2021:effective} adopt non-norm-bounded attacks using established filters to generate adversarial examples, narrowing the broader applicability of their findings.
Furthermore, Liu~\etal~\cite{liu:2024:effective} propose a novel approach for generating adversarial examples in NLP tasks with custom objectives which, however, suffers from limited applicability to the well-established classical norm-bounded attacks.

\section{Adversarial Hypervolume}
This section introduces the concept of adversarial hypervolume, a novel metric DL model robustness evaluation across varying perturbation intensities.
Initially, we outline the multi-objective adversarial problem foundational to the adversarial hypervolume concept in Section~\ref{subsec:multiobj_adv}.
Subsequently, we describe the adversarial frontier in Section~\ref{subsec:adv_frontier} and detail the methodology for computing the adversarial hypervolume in Section~\ref{subsec:adv_hv}.
We then explore the relationship between adversarial hypervolume, adversarial accuracy, and other established metrics in Section~\ref{subsec:adv_hv_robustness}.

\subsection{Multiobjecive Adversarial Problem}\label{subsec:multiobj_adv}
Consider a classifier $f: \mathcal{X}\to \mathcal{Y}$, with $\mathcal{X}\subseteq [0,1]^d$ denoting the input space, and $\mathcal{Y}\subseteq [0,1]^m$ representing the output space, which signifies a probability distribution over $m$ class labels.

\bheading{Typical Adversarial Problem.}
Adversarial attacks typically aim to determine the perturbation $\delta$, subjected to the constraint $\|\delta\|_p<\varepsilon$, where $\varepsilon$ is the perturbation budget, that can lead to a model’s misclassification. This has led to the formulation of the optimization problem below:
\begin{equation}
    \min_{\delta} \mathcal{L}_{\text{conf}}(f,x,y,\delta), \quad \text{s.t. } \|\delta\|_p \leq \varepsilon,
\end{equation}
where $\mathcal{L}_{\text{conf}}$ is the confidence loss and $p$ represents the norm used.
As misclassification inherently involves a non-differentiable target, the confidence loss is utilized as a surrogate to estimate misclassification error.
For the purposes of this paper, we adopt the minimization perspective, wherein \emph{lower} $\mathcal{L}_{\text{conf}}$ indicate \emph{higher} misclassification likelihood.

\bheading{Multiobjecive Adversarial Problem.}
In this paper, we address a multiobjective adversarial problem that incorporates two objectives: the confidence loss and the distance loss. It is formulated as follows:
\begin{equation}\label{eq:adv_problem}
    \begin{aligned}
        \min_{\delta} \mathcal{L}(f,x,y,\delta) & = (\mathcal{L}_{\text{conf}}(f,x,y,\delta),\mathcal{L}_{\text{dist}}(\delta)), \\
        \mathcal{L}_{\text{dist}}(\delta)       & = \|\delta\|_p.
    \end{aligned}
\end{equation}
This formulation delineates our goal to concurrently minimize these losses, which is more representative of real-world scenarios than the single-objective adversarial problem.

\bheading{Marginal Confidence Loss.}
The confidence loss here, as mentioned above, is a surrogate of the misclassification error. We adopt the marginal confidence loss as a confidence proxy, defined as:

\begin{equation}\label{eq:margin}
    \begin{aligned}
        \mathcal{L}_{\text{MAR\_conf}}(f,x,y,\delta) & = \left\{\begin{array}{cc}
                                                                    \text{MAR} & \text{if } \text{MAR} >0 \\
                                                                    0          & \text{otherwise}
                                                                \end{array}\right. \\
        \text{MAR}(f,x,y,\delta)                     & = f_y(x+\delta) - \max_{i\neq y}f_i(x+\delta)
    \end{aligned}
\end{equation}

We have set the non-positive margin to be zero to omit the negative values, which are not indicative of misclassification. Other confidence loss functions can be used under our framework.



\subsection{Adversarial Frontier}\label{subsec:adv_frontier}
Drawing on the multiobjective adversarial framework outlined in Section~\ref{subsec:multiobj_adv}, we incorporate the multiobjective optimization principles to determine optimal trade-off solutions.
We begin by presenting the concepts of Pareto dominance and Pareto optimality, which are central to multi-objective optimization and represent the most efficient trade-off solutions.
Subsequently, we employ an example to demonstrate the adversarial frontier delineated by these principles and to showcase the adversarial frontiers of various defensive models.

\bheading{Pareto Dominance.}
Consider two candidate perturbations $\delta^a$ and $\delta^b$. $\delta^a$ is said to \emph{dominate} $\delta^b$ if and only if, for all criteria $i$ within the set $\left\{\text{conf},\text{dist}\right\}$, $\mathcal{L}_i(\delta^a)$ is less than or equal to $\mathcal{L}_i(\delta^b)$, and there exists at least one criterion $j$ for which $\mathcal{L}_j(\delta^a)$ is strictly less than $\mathcal{L}_j(\delta^b)$.

\bheading{Pareto Optimal.}
A perturbation $\delta^*$ is said to be \emph{Pareto optimal} if and only if there is no alternative perturbation $\delta$ exists that dominates $\delta^*$. The set of all such Pareto optimal perturbations constitutes the \emph{Pareto set} in the sample space and the \emph{Pareto front} in the loss space. Within the specific context of adversarial attacks, this ensemble is termed the \emph{adversarial frontier}.

\begin{figure}[t]
    \centering
    \includegraphics[width=0.42\textwidth]{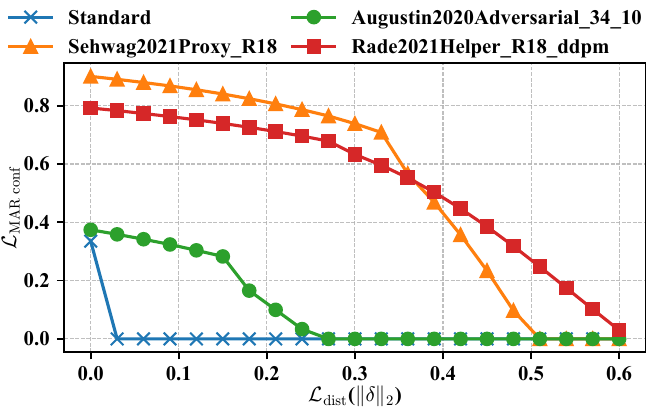}
    \vspace{-0.75\baselineskip}
    \caption{\textbf{An example of adversarial frontiers.} The adversarial frontiers of different models computed using a randomly selected image from CIFAR-10 test set. R is for ResNet and WR is for WideResNet. The label represents the name of the model in the RobustBench library.\label{fig:front}}
    \vspace{-1\baselineskip}
\end{figure}


Upon resolving the optimization problem in Equation~\eqref{eq:adv_problem} implemented with the marginal confidence loss in Equation~\eqref{eq:margin} and $\ell_2$ norm, we obtain the adversarial frontiers corresponding to various models\footnote{Here, we simply optimize the confidence loss using established attacks (PGD) with multiple fixed perturbations to derive these solutions, which is considered a basic approach to addressing the multiobjective problem (point-by-point).}. Specifically, we examine three adversarially trained models on the CIFAR-10 dataset~\cite{CIFAR10}: ResNet-18 (R-18)~\cite{rade:2021:helperbased,sehwag:2022:robust}, and WideResNet-34-10 (WR-34-10)~\cite{augustin:2020:adversarial}, all sourced from the RobustBench repository~\cite{croce:2021:robustbench}. Each model has been trained with an $\ell_2$ perturbation budget of $\epsilon=0.5$. For contrastive analysis, we incorporate a standardly trained WideResNet-28-10 model. We plot the adversarial frontiers using a randomly selected image from the test set in \figref{fig:front}.

\figref{fig:front} illustrates that the normalized confidence loss of the standardly trained model decreases significantly with minimal perturbation, in contrast with the adversarially trained counterparts that preserve the confidence level within a small perturbation margin.
Nonetheless, it is inadequate to rely solely on the adversarial accuracy of a certain perturbation level to compare the robustness of models (\eg, R-18~\cite{rade:2021:helperbased,sehwag:2022:robust}).
To address this, we introduce the adversarial hypervolume indicator, a new metric designed to quantify the robustness of models uniformly across different perturbation levels.

\subsection{Adversarial Hypervolume}\label{subsec:adv_hv}
\begin{figure}[t]
    \centering
    \includegraphics[width=0.37\textwidth]{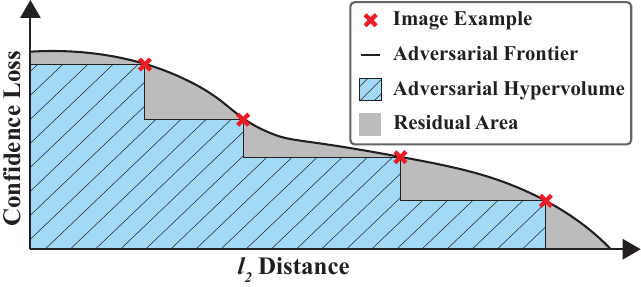}
    \vspace{-0.75\baselineskip}
    \caption{An illustration of the adversarial hypervolume. Intuitively, it functions as an approximation of the integral of confidence values, where the gray area represents the approximation's residual error.\label{fig:adv_hv_compute}}
    \vspace{-1\baselineskip}
\end{figure}

We introduce the concept of adversarial hypervolume as an approximate of the integral over the adversarial frontier. Consider a set $\Delta$ consisting of $K$ points $\Delta = \left\{\delta_0, \ldots, \delta_K\right\}$, where each $\delta_i$ satisfies $\delta_i < \epsilon$. The hypervolume indicator is computed as follows:

\begin{equation}
    HV(\Delta,\boldsymbol{r}) = \Lambda(\bigcup\limits_{\delta\in \Delta}\left\{\boldsymbol{z}\in \mathbb{R}^m | \boldsymbol{r}\preceq \boldsymbol{z} \preceq \delta\right\}).
\end{equation}
Here $\boldsymbol{r} \in \mathbb{R}^m$ represents the reference point, $\Lambda$ represents the Lebesgue measure defined as $\Lambda(Z) = \int_{\boldsymbol{z}\in Z} \mathbbm{1}_{Z}(\boldsymbol{z})d\boldsymbol{z}$, and $\mathbbm{1}_{Z}$ is the indicator function that is 1 if $\boldsymbol{z} \in Z$ and 0 otherwise. For our purposes, the reference point is set at $(0,0)$, and $m$ equals two, reflecting the consideration of two objectives. \figref{fig:adv_hv_compute} provides an illustration of the adversarial hypervolume.

\subsection{Convergence Analysis}\label{subsec:adv_hv_robustness}
In this section, we explored to measure an averaged version of robustness by integrating the adversarial frontier curve across the full spectrum of possible perturbations. While this integral is a direct representation of robustness, it is challenging to compute in practice as it requires the adversarial frontier curve to be known. Thus, we approximate the integral using a set of discrete points on the adversarial frontier curve, presenting the adversarial hypervolume as an approximation of the integral. We provide a convergence analysis of the error term between the adversarial hypervolume and the integral of the adversarial frontier in this section.



\bheading{Adversarial Frontier Curve.}
Addressing the optimization challenge delineated by Equation~\eqref{eq:adv_problem}, we derive the adversarial frontiers corresponding to various models. These frontiers are represented as a parameterized curve, denoted as $F(z)$, where $z$ indicates the perturbation magnitude. The curve characterizes the minimum $\mathcal{L}_{\text{MAR\_conf}}$ at each perturbation level:
\begin{equation}
    F(z) = \min_{\delta} \mathcal{L}_{\text{MAR\_conf}}(f,x,y,\delta), \quad \text{s.t. } \|\delta\|=z.
\end{equation}

\bheading{Integral of Adversarial Frontier.}
Averaged version of robustness can be obtained by integrating the adversarial frontier curve across the full spectrum of possible perturbations. This integral is represented as:
\begin{equation}
    \text{AH}_\text{MAR}(f,x,y,\epsilon) = \int_0^ \epsilon F(z) dz - \mathbf{e},
\end{equation}
where $\mathbf{e}$ denotes the error term that quantifies the discrepancy between the integral and the computed adversarial hypervolume.


\bheading{Approximation Error.}
For an image $(x,y)$, a classifier $f$, its corresponding adversarial frontier as $\mathcal{L}_\text{conf} = F(z)$, where $z$ is the magnitude of the perturbation, usually measured by the $p$-norm. It must be a monotonic non-increasing function by definition of the adversarial front. Moreover, we assume that the following conditions hold:
\begin{assumption}
    $F(z)$ is Lipschitz-continuous, i.e., $\exists L_0>0, \forall z_1, z_2, |F(z_1)-F(z_2)|\leq L_0|z_1-z_2|$.
\end{assumption}

When approximating the integral of the adversarial frontier over the interval $[0, \epsilon]$ using $N$ points $\left\{0, \frac{\epsilon}{N},\frac{2\epsilon}{N}, \ldots, \frac{\epsilon}{N}\right\}$, let us consider the error $\mathbf{e}_i$ within each subinterval $\left[\frac{i\epsilon}{N},\frac{(i+1)\epsilon}{N}\right]$:
\begin{equation}
    \begin{aligned}
        \mathbf{e}_i & = \int_{\frac{i\epsilon}{N}}^{\frac{(i+1)\epsilon}{N}}F(z)dz - F(\frac{(i+1)\epsilon}{N})\frac{\epsilon}{N} \\
                     & = \int_{\frac{i\epsilon}{N}}^{\frac{(i+1)\epsilon}{N}}F(z) - F(\frac{(i+1)\epsilon}{N})dz                   \\
                     & \leq \int_{\frac{i\epsilon}{N}}^{\frac{(i+1)\epsilon}{N}}L_0(z-\frac{(i+1)\epsilon}{N})dz                   \\
                     & = \frac{\epsilon^2}{N^2}L_0                                                                                 \\
    \end{aligned}
\end{equation}
where $\mathbf{e}_i$ is the error for a given subinterval. Summation of errors across all $N$ subintervals yields the total approximation error $e$, which can be expressed as follows:
\begin{equation}\label{eq:error}
    \mathbf{e} = N \mathbf{e}_i \leq \frac{\epsilon^2}{N} L_0
\end{equation}

\section{Algorithm}
This section delineates algorithms designed to identify adversarial frontiers for adversarial training approach that seeks to augment model robustness over a range of perturbation levels. Furthermore, this section presents the implementation details and analyzes the complexity inherent in these algorithms.

\subsection{Adversarial Hypervolume Computation}
The computation of adversarial hypervolume involves two primary steps: \textit{1)} identifying the adversarial frontier and \textit{2)} calculating the hypervolume of this frontier.
Specifically, identification of the adversarial frontier necessitates employing strong attacks to discover a sequence of instances with the lowest possible confidence score across different levels of perturbation. Subsequently, the adversarial hypervolume is computed using methodologies from the field of multiobjective optimization.

\bheading{Identifying the Adversarial Frontier.}
The detailed algorithm to calculate adversarial hypervolume is presented in \algref{alg:compute}, centering on the identification of the examples that constitute the frontier.
With a fixed perturbation budget of $\epsilon$ and an intent to depict the adversarial frontier using $N$ examples, we employ a strong attack method, such as the PGD attack, with perturbations $\left\{0,\frac{\epsilon}{N},\ldots,\epsilon\right\}$.
This approach results in a set of examples, each characterized by the lowest confidence score at the respective perturbation level.
Notably, examples with negative normalized scores are excluded since they represent misclassified examples, which could distort the model's average robustness assessment.
\begin{algorithm}[t]
    \caption{Adversarial Hypervolume Computation}
    \label{alg:compute}
    \begin{algorithmic}[1]
        \STATE {\bfseries Input:} Data point $(x,y)$, perturbation budget $\epsilon$, model $f$, evaluation points $N$
        \STATE {\bfseries Output:} Adversarial Hypervolume AH
        \STATE $\text{AF} = \emptyset$
        \FOR{$i=1$ {\bfseries to} $N$}
        \STATE $\delta = \text{PGD}(f,x,y,\frac{\epsilon}{N})$
        \IF{$\mathcal{L}_\text{conf}(f,x,y,\delta) < 0$}
        \STATE $\text{AF} = \text{AF} \cup \{(\frac{i}{N},0)\}$
        \STATE break
        \ELSE
        \STATE $\text{AF} = \text{AF} \cup \{(\frac{i}{N},\mathcal{L}_\text{conf}(f,x,y,\delta))\}$
        \ENDIF
        \ENDFOR
        \RETURN $\text{AH} = \text{Compute\_HV}(\text{AF})$
    \end{algorithmic}
\end{algorithm}

\bheading{Calculating the Hypervolume.}
To accurately calculate the hypervolume enclosed by the adversarial frontier, established algorithms from multiobjective optimization literature can be employed~\cite{nowak:2014:empirical,while:2012:fast}.
Specifically, in the case of a two-objective problem, the adversarial frontier is first organized in ascending order according to the perturbation level values.
Subsequently, the volume within each interval $[\frac{i\epsilon}{N}, \frac{(i+1)\epsilon}{N}]$ is determined by calculating the product of the interval's length and the maximum objective value at the end of the interval.


\bheading{Batched Adversarial Hypervolume.}
Implementing batch processing significantly enhances the computational efficiency of neural networks.
In particular, the PGD algorithm can simultaneously process multiple images from the dataset concurrently, which does not impact the optimal solutions since the aggregate of their loss values does not affect the back-propagation process.
Subsequent operations, such as sorting and hypervolume calculation for each image, can be efficiently executed in parallel using multithreading on the CPU.


\subsection{Adversarial Hypervolume Training}

In the context of adversarial hypervolume training, our algorithm harnesses examples from the adversarial frontier to enhance the training process of DL model. The adopted training algorithm unfolds as follows:

\bheading{Base Training Framework.}
Our adopted training framework integrates the TRADES framework~\cite{zhang:2019:theoretically} to serve as the cornerstone of our methodology. Employing this framework, along with a perturbation level, enables the generation of adversarial examples that augment the training process.

\bheading{Ascending Perturbation Levels.}
By progressively elevating the perturbation levels, we replicate adversarial examples stemming from the identified adversarial frontier, in accordance with the aforementioned algorithm. This method has demonstrated efficacy in enhancing model robustness and offers greater stability than conventional adversarial training approaches~\cite{addepalli:2022:efficient}.


\begin{algorithm}[t]
    \caption{Adversarial Hypervolume Training (AH Training)}
    \label{alg:adv_training}
    \begin{algorithmic}[1]
        \STATE {\bfseries Input:} network $f$ parameterized by $\theta$, dataset $\mathcal{D}$, total number of epochs $T$, attack epsilon step $\eta_1$, attack steps $K$
        \STATE {\bfseries Output:} Robust network $f_\theta$
        \STATE Randomly initialize network $f_\theta$
        \FOR{$t=1$ {\bfseries to} $T$}
        \STATE $\epsilon_t \leftarrow t \cdot \epsilon / T$
        \FOR{$i=1$ {\bfseries to} $|\mathcal{D}|$}
        \FOR{$k=1$ {\bfseries to} $K$}
        \STATE $x'_i \leftarrow \prod_{\mathbb{B}(x_i,\epsilon_t)}(x_i + \eta_1 \nabla \mathcal{L}_{\text{conf}}(f, x_i, y_i, \epsilon)) $
        \ENDFOR
        \ENDFOR
        \STATE $\mathcal{D}' = \left\{x'_1, \ldots, x'_m\right\}$
        \STATE $f_\theta \leftarrow \text{Train\_TRADES}(f_\theta, \mathcal{D} \cup \mathcal{D}')$
        \ENDFOR
    \end{algorithmic}
\end{algorithm}

\subsection{Convergence and Complexity}
The convergence of the adversarial hypervolume computation algorithm primarily hinges on the precise identification of the adversarial frontier.
Despite the challenges posed by the multiobjective nature of the problem when attempting to verify algorithmic convergence, such convergence can be substantiated by examining the optimality characteristics inherent in the PGD algorithm.
We introduce the following assumption:

\begin{assumption}
    The PGD algorithm can identify the optimal solutions with the minimal confidence loss across various perturbation budgets~\cite{madry:2018:towards}.
\end{assumption}

Under this assumption, no solutions can have an identical perturbation level, denoted as $\delta$, while presenting a lower confidence score. This implies that no solutions can dominate those found by the PGD algorithm. Consequently, the algorithm is assured to converge to the adversarial frontier.

\section{Evaluation}

\begin{table*}[t]
    \centering
    \caption{\textbf{Overall Results.} Evaluation results of standard and adversarially trained models on the CIFAR-10 dataset~\cite{CIFAR10}. Best results within the same model structure are in bold, and best overall results are highlighted with gray background. \label{tab:overall_res}}
    \vspace{-0.5\baselineskip}
    \resizebox{0.98\textwidth}{!}{
        \begin{tabular}{cllcccccc}
            \toprule
            \multirow[c]{3}{*}{\textbf{ID}} & \multirow[c]{3}{*}{\textbf{Model Structure}} & \multirow[c]{3}{*}{\textbf{Defense Method}}            & \multicolumn{3}{c}{\textbf{APGD-MAR-$\ell_2$ Attack}} & \multicolumn{3}{c}{\textbf{APGD-MAR-$\ell_\infty$ Attack}}                                                                                                                                                                                                                 \\
            \cmidrule(lr){4-6}                  \cmidrule(lr){7-9}
                                            &                                              &                                                        & \multicolumn{2}{c}{\textbf{Accuracy (\%)}}            & \multirow[c]{2}{*}{\textbf{AH ($\text{mean}_{\text{std}}$)}} & \multicolumn{2}{c}{\textbf{Accuracy (\%)}}       & \multirow[c]{2}{*}{\textbf{AH ($\text{mean}_{\text{std}}$)}}                                                                                             \\
            \cmidrule(lr){4-5}\cmidrule(lr){7-8}
                                            &                                              &                                                        & \textbf{Clean}                                        & \textbf{Robust}                                              &                                                  & \textbf{Clean}                                               & \textbf{Robust}                                                                           \\
            \midrule
            0                               & WideResNet-28-10                             & None                                                   & $\mathbf{94.78}$                                      & $\mathbf{00.35}$                                             & $\mathbf{0.1688_{0.1636}}$                       & \cellcolor{light-gray}$\mathbf{94.78}$                       & $\mathbf{00.00}$                       & $\mathbf{0.0611_{0.0864}}$                       \\
            \midrule
            1                               & ResNet-18                                    & \textbf{Ours}                                          & $\mathbf{93.50}$                                      & $76.28$                                                      & \cellcolor{light-gray}$\mathbf{0.7250_{0.3636}}$ & $\mathbf{90.15}$                                             & $54.70$                                & $\mathbf{0.4622_{0.3805}}$                       \\
            2                               & ResNet-18                                    & Rade~\etal (2022)~\cite{rade:2022:reducing}            & $90.57$                                               & $\mathbf{76.34}$                                             & $0.5889_{0.3441}$                                & $86.86$                                                      & $\mathbf{57.71}$                       & $0.3641_{0.3201}$                                \\
            3                               & ResNet-18                                    & Sehwag~\etal (2022)~\cite{sehwag:2022:robust}          & $89.76$                                               & $74.90$                                                      & $0.6382_{0.3860}$                                & $84.59$                                                      & $57.13$                                & $0.4292_{0.3745}$                                \\
            \midrule
            4                               & ResNet-50                                    & Augustin~\etal (2020)~\cite{augustin:2020:adversarial} & $\mathbf{91.08}$                                      & $\mathbf{74.08}$                                             & $\mathbf{0.5482_{0.3706}}$                       & -                                                            &                                        & -                                                \\
            \midrule
            5                               & WideResNet-28-10                             & Wang~\etal (2023)~\cite{wang:2023:better}              & \cellcolor{light-gray}$\mathbf{95.16}$                & \cellcolor{light-gray}$\mathbf{83.84}$                       & $\mathbf{0.6141_{0.2954}}$                       & $\mathbf{92.44}$                                             & \cellcolor{light-gray}$\mathbf{72.35}$ & $\mathbf{0.4286_{0.2970}}$                       \\
            5                               & WideResNet-28-10                             & Rebuffi~\etal (2021)~\cite{rebuffi:2021:fixing}        & $91.79$                                               & $79.01$                                                      & $0.5773_{0.3357}$                                & $87.33$                                                      & $62.37$                                & $0.3936_{0.3300}$                                \\
            \midrule
            6                               & WideResNet-34-10                             & Sehwag~\etal (2022)~\cite{sehwag:2022:robust}          & $90.93$                                               & $\mathbf{77.69}$                                             & $\mathbf{0.6708_{0.3812}}$                       & $86.68$                                                      & $61.77$                                & \cellcolor{light-gray}$\mathbf{0.4993_{0.3972}}$ \\
            7                               & WideResNet-34-10                             & Augustin~\etal (2020)~\cite{augustin:2020:adversarial} & $\mathbf{92.23}$                                      & $76.75$                                                      & $0.5691_{0.3654}$                                & -                                                            &                                        & -                                                \\
            8                               & WideResNet-34-10                             & Huang~\etal (2021)~\cite{huang:2021:exploring}         &                                                       & -                                                            & -                                                & $90.56$                                                      & $63.55$                                & $0.4992_{0.3608}$                                \\
            9                               & WideResNet-34-10                             & Huang~\etal (2021)~\cite{huang:2021:exploring} (EMA)   &                                                       & -                                                            & -                                                & $\mathbf{91.23}$                                             & $\mathbf{64.57}$                       & $0.4978_{0.3605}$                                \\
            \midrule
            10                              & WideResNet-70-16                             & Rebuffi~\etal (2021)~\cite{rebuffi:2021:fixing}        & $\mathbf{92.41}$                                      & $\mathbf{80.56}$                                             & $\mathbf{0.6165_{0.3376}}$                       & $\mathbf{88.54}$                                             & $\mathbf{66.04}$                       & $\mathbf{0.4395_{0.3415}}$                       \\

            \bottomrule
        \end{tabular}
    }
\end{table*}

\begin{table*}
    \centering
    \caption{\textbf{Wilcoxon rank-sum test.} The best results are marked in bold and gray background. The mean value is reported, the mark ($+/-/=$, better, worse, and equal) and the $p$-value are shown in the parentheses.\label{tab:wilcoxon}}
    \vspace{-0.5\baselineskip}
    \begin{tabular}[c]{lcccc}
        \toprule
        \multirow[c]{2}{*}{\textbf{Defense Method}}   & \multicolumn{2}{c}{\textbf{APGD-MAR-$\ell_2$ Attack}} & \multicolumn{2}{c}{\textbf{APGD-MAR-$\ell_\infty$ Attack}}                                                                                               \\
        \cmidrule(lr){2-3}\cmidrule(lr){4-5}
                                                      & \textbf{Robust Acc.}                                  & \textbf{AH}                                                & \textbf{Robust Acc.}                              & \textbf{AH}                             \\
        \midrule
        Ours (Mean Value)                             & \cellcolor{light-gray}$\mathbf{76.35}$                & \cellcolor{light-gray}$\mathbf{0.7221}$                    & $55.36$                                           & \cellcolor{light-gray}$\mathbf{0.4626}$ \\
        Sehwag~\etal (2022)~\cite{sehwag:2022:robust} & $74.43(-, 0.0001)$                                    & $0.6371(-, 0.0001)$                                        & $57.32(+, 0.0002)$                                & $0.4292(-, 0.0001)$                     \\
        Rade~\etal (2022)~\cite{rade:2022:reducing}   & $76.34(=, 0.5932)$                                    & $0.5891(-, 0.0001)$                                        & \cellcolor{light-gray}$\mathbf{57.47(+, 0.0002)}$ & $0.3639(-, 0.0001)$                     \\
        \bottomrule
    \end{tabular}

\end{table*}

\subsection{Experimental Setup}
\bheading{Datasets and Models.}
To evaluate adversarial robustness comprehensively, we calculate the adversarial hypervolume of models equipped with various defensive strategies and adversarial training techniques on both popular benchmark dataset CIFAR-10 dataset~\cite{CIFAR10} and natural image dataset, ImageNet dataset.~\cite{imagenet_cvpr09}.

\lheading{Datasets.}
For CIFAR-10 dataset~\cite{CIFAR10}, it consists of $50,000$ training images and $10,000$ testing images, each with dimensions of $32 \times 32$ pixels in full color, across ten distinct categories.
This dataset is recognized as a pivotal benchmark for adversarial robustness evaluation. Additionally, our training includes semi-supervised dataset generated by generative models from Wang~\etal's work~\cite{wang:2023:better}.
For ImageNet dataset~\cite{imagenet_cvpr09}, its test set contains $100,000$ images from $1,000$ objective classes. We utilize it for providing references of our proposed metric.

\lheading{Models.}
We utilize models from RobustBench~\cite{croce:2021:robustbench}, which encompasses both standardly and adversarially trained models from SOTA works in the field of adversarial robustness~\cite{wang:2023:better,rebuffi:2021:fixing,sehwag:2022:robust,rade:2022:reducing,augustin:2020:adversarial,huang:2021:exploring,wong:2020:fast,salman:2020:do}.

\bheading{Attack for Evaluation.}
To conduct the evaluation, a modified version of the APGD attack, named APGD-MAR, was utilized within the AutoAttack framework~\cite{croce:2020:reliable}.
The APGD-MAR, an adaptation of the APGD-CE attack, is specifically designed to identify adversarial examples by employing a marginal loss function, which is a fundamental component of our metric's implementation.
Empirical evidence has demonstrated that employing solely projected gradient-based attacks is effective for assessing adversarial robustness, as observed by the negligible difference in reported adversarial accuracy when compared to the Square Attack~\cite{croce:2020:reliable}.

Attack parameters were standardized across evaluations, with the number of iterations fixed at $20$ and the maximum perturbation magnitude at $8/255$ for the $\ell_\infty$ attack and $0.5$ for the $\ell_2$ attack on the CIFAR-10 dataset.
For the ImageNet dataset,  these parameters were adjusted to $20$ steps and a maximum perturbation of $4/255$ for the $\ell_\infty$ attack, , in line with the attack constraints in RobustBench~\cite{croce:2021:robustbench}.
Furthermore, the perturbation magnitude was divided into $N=10$ equal intervals to facilitate the computation of the adversarial hypervolume.

\bheading{Defense Methods.}
We consider measuring the robustness of the models under both methods to improve the inherent robustness (adversarial training) and methods to protect the input of the model (input transformation).

\lheading{Adversarial Training.}
Adversarial training enhances the model robustness by utilizing augmented data samples crafted using various attack strategies. In this work, we use pre-trained adversarially trained models using SOTA training methods, including \cite{madry:2018:towards,wang:2023:better,rebuffi:2021:fixing,sehwag:2022:robust,rade:2022:reducing,augustin:2020:adversarial,huang:2021:exploring,robustness}.
These models rank highly on the RobustBench leaderboard~\cite{croce:2021:robustbench} and are widely used as strong robust baselines.

\lheading{Input Transformation.}
Input transformations are straightforward defensive mechanisms that can be easily integrated into systems with adversarially trained models, contributing to an improvement in overall robustness.
Our research reexamines several of these techniques, JPEG compression~\cite{guo:2018:countering}, Feature Squeezing, and Spatial Smoothing~\cite{xu:2018:feature}. Notably, they are known to be weak and breakable defenses~\cite{athalye:2018:obfuscated} and adversarial accuracy may not be adequate for comparing their efficacy.

\subsection{Overall Results} \label{subsec:overall}
The clean accuracy, adversarial accuracy, and adversarial hypervolume of the models trained and evaluated under both $\ell_2$ and $\ell_\infty$ threat models on the CIFAR-10 dataset are presented in \tabref{tab:overall_res}.
Specially, the mean and the standard deviation of the adversarial hypervolume metrics calculated on the test set are reported.
Generally, the adversarial accuracy and adversarial hypervolume of the models are consistent, with the adversarial hypervolume providing a more comprehensive evaluation of the model's robustness.
In the following sections, we analyze the results and discuss the implications of the findings.

\bheading{Effectiveness of AH Training.}
The AH training algorithm has demonstrated substantial efficacy in augmenting the adversarial robustness of machine learning models, as evidenced by the improved adversarial accuracy and hypervolume metrics.  Specifically, the models utilizing our training algorithm achieved the best adversarial hypervolume among the models trained with the same structure under both $\ell_2$ and $\ell_\infty$ attacks. Moreover, our models reported comparable adversarial accuracy to the SOTA models out of all model structures, demonstrating the effectiveness of our training algorithm.

\bheading{Standardly \emph{v.s.} Adversarially Trained.}
The adversarial hypervolume of the standardly trained models is significantly lower than that of the adversarially trained models as shown in \tabref{tab:overall_res}. As the initial confidence score of the standardly trained models is usually high, we can conclude that their confidence score decreases significantly when exposed to adversarial examples with minimal perturbation, leading to a lower adversarial hypervolume. It is evidenced by the example in \figref{fig:front}.

\bheading{AH for Model Selection.}
When two models have similar adversarial accuracy, adversarial hypervolume serves as an additional metric to differentiate their robustness, as shown in the comparison between our model and the model trained by Rade~\etal~\cite{rade:2022:reducing}. Adversarial hypervolume provides a more comprehensive evaluation of the model's robustness, which is crucial for selecting the most robust model for deployment.

\subsection{Further Validation of Effectiveness}
To further validate the effectiveness of our method and diminish the potential impact of randomness, we conducted a Wilcoxon rank-sum test on our model and the model trained by Rade~\etal~\cite{rade:2022:reducing} and Sehwag~\etal (2022)~\cite{sehwag:2022:robust} under $\ell_2$ and $\ell_\infty$ attacks.
Specifically, we adopted the same training and evaluation settings for our model in \secref{subsec:overall} and computed the adversarial accuracies and adversarial hypervolumes of the models under $10$ different random seeds.
The two baseline models are from from RobustBench~\cite{croce:2021:robustbench}.
For fair comparison, both the baseline models and our model are ResNet-18 models.

\bheading{Results}
The results of the Wilcoxon rank-sum test are shown in Table~\ref{tab:wilcoxon}. The table contains the results robust accuracy and adversarial hypervolume under $\ell_2$ and $\ell_\infty$ attacks, with the comparison results determined with the significance level of $p=0.05$.
Our methods achieved the highest adversarial hypervolume under both attacks, with the highest robust accuracy under the $\ell_2$ attack.
For the robust accuracy under the $\ell_\infty$ attack, our model achieved comparable performance with only $2\%$ lower than the best model.
The Wilcoxon rank-sum test results further confirm the effectiveness of our method in enhancing the adversarial robustness of the models.

\begin{table*}[t]
    \centering
    \caption{\textbf{Ablation Study.} Ablation study for our training algorithm under CIFAR-10 dataset.\label{tab:ablation_study}}
    \vspace{-0.5\baselineskip}
    \begin{tabular}{lcccccc}
        \toprule
        \multirow[c]{3}{*}{\textbf{Defense Method}} & \multicolumn{3}{c}{\textbf{APGD-MAR-$\ell_2$ Attack}} & \multicolumn{3}{c}{\textbf{APGD-MAR-$\ell_\infty$ Attack}}                                                                                                                                                                                                                 \\
        \cmidrule(lr){2-4}                  \cmidrule(lr){5-7}
                                                    & \multicolumn{2}{c}{\textbf{Accuracy (\%)}}            & \multirow[c]{2}{*}{\textbf{AH ($\text{mean}_{\text{std}}$)}} & \multicolumn{2}{c}{\textbf{Accuracy (\%)}}       & \multirow[c]{2}{*}{\textbf{AH ($\text{mean}_{\text{std}}$)}}                                                                                             \\
        \cmidrule(lr){2-3}\cmidrule(lr){5-6}
                                                    & \textbf{Clean}                                        & \textbf{Robust}                                              &                                                  & \textbf{Clean}                                               & \textbf{Robust}                                                                           \\
        \midrule
        None                                        & \cellcolor{light-gray}$\mathbf{93.50}$                & \cellcolor{light-gray}$\mathbf{76.28}$                       & $0.7250_{0.3636}$                                & \cellcolor{light-gray}$\mathbf{90.15}$                       & \cellcolor{light-gray}$\mathbf{54.70}$ & $0.4622_{0.3805}$                                \\
        w./o. ascending                             & $93.59$                                               & $74.12$                                                      & $0.6978_{0.3713}$                                & $89.84$                                                      & $53.03$                                & $0.4556_{0.3809}$                                \\
        w./o. data                                  & $90.49$                                               & $63.74$                                                      & \cellcolor{light-gray}$\mathbf{0.7292_{0.3765}}$ & $86.60$                                                      & $40.84$                                & \cellcolor{light-gray}$\mathbf{0.5626_{0.4019}}$ \\
        \bottomrule
    \end{tabular}

\end{table*}

\begin{table}
    \centering
    \caption{\textbf{Input Transformation.} Results on Input Transformation Defense Methods. Clean is for clean accuracy and Robust is for robust accuracy. The best results are marked in bold and gray background.\label{tab:input}}
    \vspace{-0.5\baselineskip}
    \resizebox{0.48\textwidth}{!}{
        \begin{tabular}[c]{lccccc}
            \toprule
            \multirow[c]{2}{*}{\textbf{Method}} & \multirow[c]{2}{*}{\textbf{Clean}}     & \multicolumn{2}{c}{\textbf{APGD-MAR-$\ell_2$ Attack}} & \multicolumn{2}{c}{\textbf{APGD-MAR-$\ell_\infty$ Attack}}                                                                                             \\
            \cmidrule(lr){3-4}\cmidrule(lr){5-6}
                                                &                                        & \textbf{Robust}                                       & \textbf{AH}                                                & \textbf{Robust}                        & \textbf{AH}                                      \\
            \midrule
            None                                & \cellcolor{light-gray}$\mathbf{94.78}$ & $00.35$                                               & $0.1688_{0.1636}$                                          & $00.00$                                & $0.0611_{0.0864}$                                \\
            JPEG~\cite{guo:2018:countering}     & $91.28$                                & $07.61$                                               & $0.0521_{0.2047}$                                          & $03.19$                                & $0.0001_{0.0098}$                                \\
            FS~\cite{xu:2018:feature}           & $94.24$                                & $03.28$                                               & $0.0418_{0.1989}$                                          & $01.61$                                & $0.0394_{0.1945}$                                \\
            SPS~\cite{xu:2018:feature}          & $87.33$                                & \cellcolor{light-gray}$\mathbf{20.21}$                & \cellcolor{light-gray}$\mathbf{0.2033_{0.3765}}$           & \cellcolor{light-gray}$\mathbf{05.91}$ & \cellcolor{light-gray}$\mathbf{0.0105_{0.0929}}$ \\
            \bottomrule
        \end{tabular}
    }

\end{table}

\subsection{Results on Input Transformation}
Input transformation techniques, while recognized as vulnerable and easily circumvented defenses, can be conveniently integrated into systems to marginally enhance robustness.
Despite its simplicity, evaluation of performance under the metric of adversarial accuracy presents challenges in making meaningful comparisons and deriving insights. Therefore, we employed the metric of adversarial hypervolume in our experimental assessment of the robustness of these methods.


\bheading{Results.}
Table \ref{tab:input} presents the results for the various input transformation methods.
All methods display significantly lower adversarial accuracies under both attack types.
Furthermore, when employing prevalent perturbation budgets, the $\ell_\infty$ attack is observed to be more potent than the $\ell_2$ attack, as evidenced by the lower adversarial accuracy and hypervolume.
Of all the input transformation methods evaluated, only spatial smoothing imparts some resistance to $\ell_2$ attack, whereas none effectively withstand $\ell_\infty$ attack.
The adversarial hypervolume metric suggests that while these transformations can resist minor noise, they are ineffective against substantial perturbations.

\subsection{Ablation Study}
The efficacy of our adversarial hypervolume training techniques was further investigated through an ablation study. The study involved two modifications:

\begin{itemize}
    \item \textbf{Without Ascending.} We remove the ascending training strategy from the adversarial hypervolume training method and use the fixed perturbation strategy as done in most adversarial training methods.
    \item \textbf{Without Synthetic data.} We remove the synthetic data generated by the diffusion model and only use the original data for training.
\end{itemize}

\bheading{Results.}
The findings, as presented in \tabref{tab:ablation_study}, indicate that the standard approach outperforms alternative strategies concerning adversarial robustness, achieving the highest performance in adversarial accuracy during both $\ell_2$ and $\ell_\infty$ attacks, with the slight expense of clean accuracy.
Elimination of the ascending technique resulted in a marginal reduction in robustness metrics.

\lheading{Mismatch.}
In \tabref{tab:ablation_study}, we noticed an unusual increase in the adversarial hypervolume when the synthetic data was removed, along with a decrease in the adversarial accuracy.
This discrepancy suggests that the model can be over-confident on its predictions when the synthetic data is removed, leading to a lower adversarial accuracy but a higher adversarial hypervolume.
This phenomenon reveals that both adversarial accuracy and adversarial hypervolume are essential when considering the evaluation of adversarial robustness.

\begin{figure}[!t]
    \centering
    \subfloat[Clean Accuracy]{\includegraphics[width=1.5in]{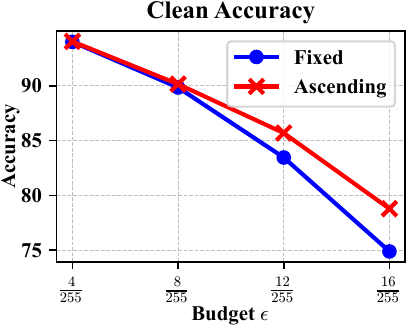}%
        \label{fig_second_case}}
    \hfil
    \subfloat[Robust Accuracy]{\includegraphics[width=1.5in]{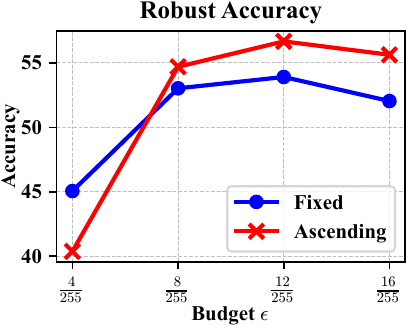}%
        \label{fig_first_case}}
    \caption{\textbf{Accuracy Comparison.} Results of the clean and robust accuracy trained under fixed budget and ascending budget strategy.\label{fig:ascending}}
\end{figure}

\bheading{Further Results on Ascending.}
The effectiveness of ascending strategy of the perturbation budget is further validated, with the results under $\ell_\infty$ attack shown in \figref{fig:ascending} and \tabref{tab:ah_compare}.
The ascending strategy significantly improves the adversarial hypervolume of the models under $\ell_\infty$ attacks, as evidenced by the higher adversarial hypervolume of the models trained with the ascending strategy compared to the models trained with the fixed budget strategy.

\begin{table}
    \centering
    \caption{\textbf{Ascending Strategy.} Adversarial hypervolume metric of models trained using different perturbation budgets.\label{tab:ah_compare}}
    \vspace{-0.5\baselineskip}
    \resizebox{0.25\textwidth}{!}{
        \begin{tabular}[c]{lcc}
            \toprule
            $\epsilon$ & None                                             & No ascending                                     \\
            \midrule
            $ 4/255$   & $0.5444_{0.3686}$                                & \cellcolor{light-gray}$\mathbf{0.5893_{0.3699}}$ \\
            $ 8/255$   & \cellcolor{light-gray}$\mathbf{0.4656_{0.3805}}$ & $0.4622_{0.3809}$                                \\
            $12/255$   & \cellcolor{light-gray}$\mathbf{0.3431_{0.3429}}$ & $0.3288_{0.3421}$                                \\
            $16/255$   & \cellcolor{light-gray}$\mathbf{0.2319_{0.2820}}$ & $0.1978_{0.2585}$                                \\
            \bottomrule
        \end{tabular}
    }
\end{table}


\subsection{Results of ImageNet Dataset}
As a further validation of our proposed metric, we conducted experiments on the ImageNet dataset to evaluate the adversarial robustness of models under the APGD-MAR-$\ell_\infty$ attack.
The results are shown in Table~\ref{tab:imagenet}.
Generally, the adversarial accuracy and adversarial hypervolume of the models are consistent, both indicating a increase in adversarial robustness with advanced defense methods and more complicated model structures.
Further investigations on the ImageNet dataset will be conducted in our future work to provide a more comprehensive understanding of the relationship between adversarial accuracy and adversarial hypervolume.

\begin{table}
    \centering
    \caption{\textbf{Results on ImageNet dataset.} Results on ImageNet dataset under APGD-MAR-$\ell_\infty$ Attack.\label{tab:imagenet}}
    \vspace{-0.5\baselineskip}
    \resizebox{0.48\textwidth}{!}{
        \begin{tabular}[c]{clccc}
            \toprule
            \multirow[c]{2}{*}{\textbf{Model Structure}} & \multirow[c]{2}{*}{\textbf{Defense Method}} & \multicolumn{2}{c}{\textbf{Accuracy}}  & \multirow[c]{2}{*}{\textbf{AH}}                                                           \\
            \cmidrule{3-4}
                                                         &                                             & \textbf{Clean}                         & \textbf{Robust}                        &                                                  \\
            \midrule
            ResNet-50                                    & Standard                                    & \cellcolor{light-gray}$\mathbf{75.89}$ & $26.98$                                & $0.3093_{0.3592}$                                \\
            ResNet-50                                    & Wong~\etal~\cite{wong:2020:fast}            & $52.71$                                & $49.55$                                & $0.3100_{0.3905}$                                \\
            ResNet-50                                    & Engstrom~\etal~\cite{robustness}            & $61.93$                                & $58.84$                                & $0.3718_{0.4038}$                                \\
            ResNet-50                                    & Salman~\etal~\cite{salman:2020:do}          & $63.32$                                & \cellcolor{light-gray}$\mathbf{60.61}$ & \cellcolor{light-gray}$\mathbf{0.3721_{0.3985}}$ \\
            ResNet-18                                    & Salman~\etal~\cite{salman:2020:do}          & $51.92$                                & $49.00$                                & $0.2424_{0.3449}$                                \\
            \bottomrule
        \end{tabular}
    }
\end{table}
\subsection{Convergence Results}
\bheading{Analysis Settings.}
We employ an adversarially trained model~\cite{madry:2018:towards} to demonstrate the convergence properties of our algorithm in calculating the adversarial hypervolume.
For these computations, we fixed $N=20$ as the upper bound for the number of points considered, and we approximated the true integral of the confidence score using these computed values.
To derive the theoretical error function, we formulate and solve an optimization problem that minimizes the squared loss between the theoretically predicted loss (as specified in Equation~(\ref{eq:error})) and the observed error measurements, with $L_0$ treated as a variable via CVXPY~\cite{diamond:2016:cvxpy}.
In additional configurations, we adopted the $\ell_2$ norm for distance loss and the normalized cross-entropy loss for the confidence metric.

\begin{figure}[t]
    \centering
    \includegraphics[width=0.38\textwidth]{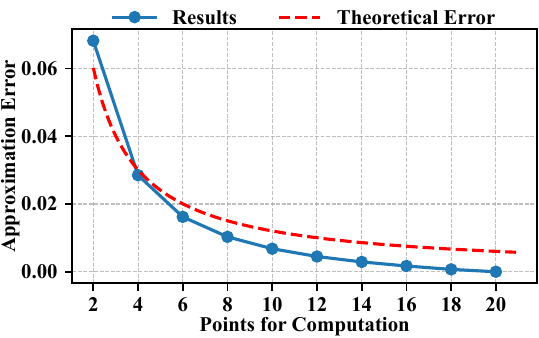}
    \vspace{-0.5\baselineskip}
    \caption{An illustration of computed and theoretical error between the adversarial hypervolume and the true integral of the confidence loss (Adversarial hypervolume approximated using $N=20$ points as a proxy). The computed values are represented by the blue line, while the theoretical error function is depicted with a red dashed line.\label{fig:hv_converge}}
\end{figure}

\bheading{Results.}
The results, illustrated in Figure \ref{fig:hv_converge}, verify that the error diminishes in direct proportion to the inverse of the number of points utilized in calculating the adversarial hypervolume.
Furthermore, once the point count surpasses 10, the error becomes negligible, suggesting that our approximation achieves sufficient precision.


\section{Discussion}




\begin{figure}[t]
    \centering
    \includegraphics[width=0.42\textwidth]{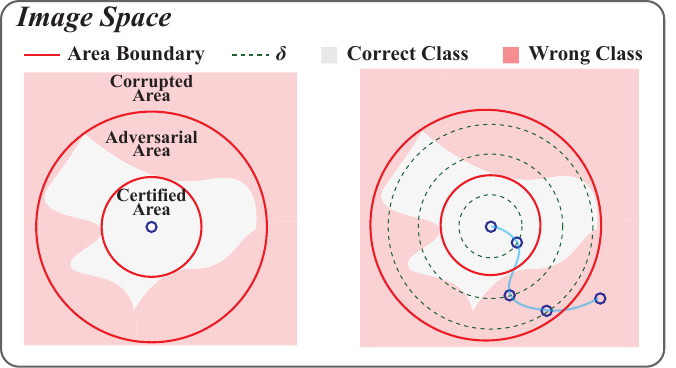}
    \caption{An illustration of three regions and the adversarial frontier in the image sample space. The blue curve denotes the adversarial frontier.\label{fig:intuition-2}}
\end{figure}

\subsection{Why Adversarial Hypervolume?}
The adoption of adversarial hypervolume as a metric to evaluate the robustness of neural network models is justified by several compelling reasons:
\begin{itemize}
    \item \textbf{Multiobjecive Nature.} Since attacking a neural network model constitutes a multiobjective optimization problem, a metric that encapsulates the trade-offs between distance and confidence is essential. Adversarial hypervolume aptly fulfills this requirement.
    \item \textbf{Limitations of Current Metrics.} Although adversarial accuracy offers a straightforward assessment of robustness, it falls short in guiding researchers to pinpoint areas of improvement and to differentiate between models with similar adversarial accuracy scores. Adversarial sparsity and probability metrics attempt to capture the robustness from different perspectives but are constrained to analyzing the misclassified sample space. In contrast, adversarial hypervolume emphasizes the correctly classified sample space.
    \item \textbf{A Natural Interpretation.} Beyond these benefits, our metric provides an intuitive interpretation since it represents the sample set with worst confidence scores across the perturbation levels. Our further research on this adversarial frontier is planned to yield insights for the development of more robust models.
\end{itemize}

\subsection{Intuition of the Proposed Metric}
This subsection introduces the conceptual foundation underlying the proposed metric. For visual representation of the image sample space, the misclassification area is highlighted in red, while a grey zone encircles the correctly classified examples adjacent to the original sample, as illustrated in \figref{fig:intuition-2}.
The space is segmented into three distinct regions relative to the $\ell_2$-norm: the certified region, the adversarial region, and the corrupted region.

\begin{itemize}
    \item \textbf{Certified Region.} Within this domain, demarcated by the established perturbation limit, all samples are guaranteed to be classified accurately. The range of this particular area partially reflects the robustness of the model.
    \item \textbf{Adversarial Region.} This zone contains a mix of correctly and incorrectly classified samples. The proportion of misclassified instances within this region correlates with the level of adversarial sparsity~\cite{oliver:2023:howmany}.
    \item \textbf{Corrupted Region.} Samples in this domain are invariably misclassified.
\end{itemize}

\figref{fig:intuition-2} presents a schematic depiction of the three regions within the image sample space, featuring the adversarial frontier as a blue curve. This frontier delineates samples with minimum confidence scores at specific perturbation intensities.

\begin{figure}[t]
    \centering
    \includegraphics[width=0.32\textwidth]{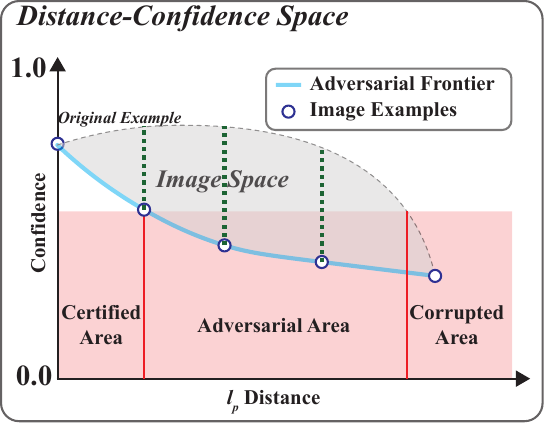}
    \caption{An illustration of the projection of the image space onto the distance-confidence space.\label{fig:intuition}}
\end{figure}

Upon projecting the aforementioned space onto a distance-confidence plane, \figref{fig:intuition} emerges, with the shaded portion exemplifying the image space across the three delineated areas. Our metric, termed adversarial hypervolume, calculates an average for the certified area while also accounting for the confidence score, thereby providing enriched information.

\subsection{Extension to Other Types of Attacks}
The adversarial hypervolume metric is computed under a bi-objective adversarial problem setting, where the perturbation budget and confidence score are the two objectives.
Therefore, the extension of the metric can be divided into two directions: \emph{new implementation} and \emph{new problem definition}.

\bheading{New Implementation.}
The adversarial hypervolume metric can be directly implemented in $\ell_0$ attacks, with the perturbation budget being the $\ell_0$ norm and the confidence score being the marginal confidence score. Recently, $\ell_0$ adversarial attacks, also known as sparse adversarial attacks, have attracted increasing attention~\cite{williams:2023:blackbox}, thereby necessitating the development of a robust metric for evaluating the adversarial robustness of models against $\ell_0$ attacks. Moreover, under this attack setting, the problem definition can be extended to a tri-objective problem by incorporating the perturbation budget as an additional objective, leading to a more comprehensive evaluation of the model's robustness. We plan to explore this direction in our future work.

\bheading{New Problem Definition.}
As new types of adversarial attacks emerge, \eg adversarial patch attacks~\cite{brown:2017:adversarial}, we can extend the adversarial hypervolume metric to accommodate these new attack settings. For instance, in adversarial patch attacks, the perturbation budget can be defined as the size of the patch, while the confidence score can be the marginal confidence score. By incorporating these two objectives, we can evaluate the robustness of models against adversarial patch attacks. We aim to explore this direction in our future work.

\subsection{Connection and Difference with Other Metrics}

\bheading{Adversarial Accuracy.}
To establish the relationship between adversarial accuracy and adversarial hypervolume, we first specify the expression for adversarial accuracy at a specific perturbation level $z$ using the sign function:
\begin{equation}
    \text{AA}(f,x,y,z) = \text{sign}_+(F(z)).
\end{equation}

Subsequently, we define the average adversarial accuracy across the entire range of perturbations as:
\begin{equation}
    \overline{\text{AA}}(x,y,\epsilon) = \int_0^\epsilon \text{sign}_+(F(z)) dz.
\end{equation}

It is evident that by employing the marginal loss as a confidence proxy, the adversarial hypervolume encompasses a broader array of information than adversarial accuracy alone.
Moreover, another implementation using the normalized confidence loss enhances this metric by integrating the insights gained from misclassified examples, rendering it a more comprehensive and robust measure of model resilience.

\bheading{Adversarial Sparsity.}
Since the adversarial sparsity~\cite{oliver:2023:howmany} quantifies the extent of the misclassified region within the entire sample space, adversarial accuracy at a given perturbation level $z$ can be expressed as:
\begin{equation}
    \text{AA}(f,x,y,z) = 1-\text{sign}_+(1-\text{AS}(f,x,y,z))
\end{equation}

The integration of adversarial accuracy over the complete range of perturbations is represented by:
\begin{equation}
    \overline{\text{AA}}(f,x,y,\epsilon) = \int_0^\epsilon 1-\text{sign}_+(1-\text{AS}(f,x,y,z))dz
\end{equation}

\bheading{Connection and Difference.}
Our metric emphasizes correctly classified instances, in contrast to adversarial sparsity, which focuses on the analysis of misclassified samples.
Consequently, it constitutes an important complement for assessing a model's robustness.
Moreover, our metric captures the average effect across the perturbation space, diverging from adversarial sparsity that is confined to a specific perturbation distance.
This difference is rooted in the need for computational viability: the calculation of adversarial sparsity over a comprehensive perturbation range is impractical due to its prohibitive computational demand. Conversely, our metric design emphasizes computational efficiency, enabling its application to large-scale datasets.

\section{Conclusion}
We have proposed the adversarial hypervolume metric, offering a comprehensive evaluation of deep learning model robustness across varying perturbation intensities. Our findings  have demonstrated that this metric surpasses traditional adversarial accuracy in capturing incremental robustness improvements and provides a deeper understanding of model resilience. The adversarial hypervolume facilitates the development of robust deep learning systems and sets a new standard for robustness assessment in the face of adversarial threats.

\bibliographystyle{IEEEtran}
\bibliography{000-paper}

\begin{thebibliography}{10}
\providecommand{\url}[1]{#1}
\csname url@samestyle\endcsname
\providecommand{\newblock}{\relax}
\providecommand{\bibinfo}[2]{#2}
\providecommand{\BIBentrySTDinterwordspacing}{\spaceskip=0pt\relax}
\providecommand{\BIBentryALTinterwordstretchfactor}{4}
\providecommand{\BIBentryALTinterwordspacing}{\spaceskip=\fontdimen2\font plus
\BIBentryALTinterwordstretchfactor\fontdimen3\font minus \fontdimen4\font\relax}
\providecommand{\BIBforeignlanguage}[2]{{%
\expandafter\ifx\csname l@#1\endcsname\relax
\typeout{** WARNING: IEEEtran.bst: No hyphenation pattern has been}%
\typeout{** loaded for the language `#1'. Using the pattern for}%
\typeout{** the default language instead.}%
\else
\language=\csname l@#1\endcsname
\fi
#2}}
\providecommand{\BIBdecl}{\relax}
\BIBdecl

\bibitem{madry:2018:towards}
A.~Madry, A.~Makelov, L.~Schmidt, D.~Tsipras, and A.~Vladu, ``Towards deep learning models resistant to adversarial attacks,'' in \emph{6th International Conference on Learning Representations, ({ICLR})}.\hskip 1em plus 0.5em minus 0.4em\relax OpenReview.net, 2018.

\bibitem{ren:2018:learning}
M.~Ren, W.~Zeng, B.~Yang, and R.~Urtasun, ``Learning to reweight examples for robust deep learning,'' in \emph{Proceedings of the 35th International Conference on Machine Learning, ({ICML})}, ser. Proceedings of Machine Learning Research.\hskip 1em plus 0.5em minus 0.4em\relax {PMLR}, 2018.

\bibitem{salman:2019:provably}
H.~Salman, J.~Li, I.~P. Razenshteyn, P.~Zhang, H.~Zhang, S.~Bubeck, and G.~Yang, ``Provably robust deep learning via adversarially trained smoothed classifiers,'' in \emph{Advances in Neural Information Processing Systems 32: Annual Conference on Neural Information Processing Systems 2019, ({NeurIPS})}, 2019.

\bibitem{dong:2020:adversarial}
Y.~Dong, Z.~Deng, T.~Pang, J.~Zhu, and H.~Su, ``Adversarial distributional training for robust deep learning,'' in \emph{Advances in Neural Information Processing Systems 33: Annual Conference on Neural Information Processing Systems 2020, ({NeurIPS})}, 2020.

\bibitem{Sadeghi:2020:taxonomy}
\BIBentryALTinterwordspacing
K.~Sadeghi, A.~Banerjee, and S.~K.~S. Gupta, ``A system-driven taxonomy of attacks and defenses in adversarial machine learning,'' \emph{{IEEE} Trans. Emerg. Top. Comput. Intell.}, vol.~4, no.~4, pp. 450--467, 2020. [Online]. Available: \url{https://doi.org/10.1109/TETCI.2020.2968933}
\BIBentrySTDinterwordspacing

\bibitem{szegedy:2014:intriguing}
C.~Szegedy, W.~Zaremba, I.~Sutskever, J.~Bruna, D.~Erhan, I.~J. Goodfellow, and R.~Fergus, ``Intriguing properties of neural networks,'' in \emph{2nd International Conference on Learning Representations ({ICLR})}, 2014.

\bibitem{goodfellow:2015:explaining}
I.~J. Goodfellow, J.~Shlens, and C.~Szegedy, ``Explaining and harnessing adversarial examples,'' in \emph{3rd International Conference on Learning Representations, ({ICLR})}, 2015.

\bibitem{kurakin:2017:adversarial}
A.~Kurakin, I.~J. Goodfellow, and S.~Bengio, ``Adversarial examples in the physical world,'' in \emph{5th International Conference on Learning Representations, {ICLR}}.\hskip 1em plus 0.5em minus 0.4em\relax OpenReview.net, 2017.

\bibitem{dong:2019:efficient}
Y.~Dong, H.~Su, B.~Wu, Z.~Li, W.~Liu, T.~Zhang, and J.~Zhu, ``Efficient decision-based black-box adversarial attacks on face recognition,'' in \emph{{IEEE} Conference on Computer Vision and Pattern Recognition, {CVPR}}.\hskip 1em plus 0.5em minus 0.4em\relax Computer Vision Foundation / {IEEE}, 2019.

\bibitem{cao:2019:adversarial}
Y.~Cao, C.~Xiao, B.~Cyr, Y.~Zhou, W.~Park, S.~Rampazzi, Q.~A. Chen, K.~Fu, and Z.~M. Mao, ``Adversarial sensor attack on lidar-based perception in autonomous driving,'' in \emph{Proceedings of the 2019 {ACM} {SIGSAC} Conference on Computer and Communications Security, ({CCS})}.\hskip 1em plus 0.5em minus 0.4em\relax {ACM}, 2019.

\bibitem{carlini:2017:towards}
N.~Carlini and D.~A. Wagner, ``Towards evaluating the robustness of neural networks,'' in \emph{2017 {IEEE} Symposium on Security and Privacy, {SP}}.\hskip 1em plus 0.5em minus 0.4em\relax {IEEE} Computer Society, 2017.

\bibitem{athalye:2018:obfuscated}
A.~Athalye, N.~Carlini, and D.~A. Wagner, ``Obfuscated gradients give a false sense of security: Circumventing defenses to adversarial examples,'' in \emph{Proceedings of the 35th International Conference on Machine Learning, {ICML}}, ser. Proceedings of Machine Learning Research.\hskip 1em plus 0.5em minus 0.4em\relax {PMLR}, 2018.

\bibitem{dezfooli:2016:deepfool}
S.~Moosavi{-}Dezfooli, A.~Fawzi, and P.~Frossard, ``Deepfool: {A} simple and accurate method to fool deep neural networks,'' in \emph{2016 {IEEE} Conference on Computer Vision and Pattern Recognition, ({CVPR})}.\hskip 1em plus 0.5em minus 0.4em\relax {IEEE} Computer Society, 2016.

\bibitem{croce:2020:reliable}
F.~Croce and M.~Hein, ``Reliable evaluation of adversarial robustness with an ensemble of diverse parameter-free attacks,'' in \emph{Proceedings of the 37th International Conference on Machine Learning, ({ICML})}, ser. Proceedings of Machine Learning Research.\hskip 1em plus 0.5em minus 0.4em\relax {PMLR}, 2020.

\bibitem{andriushchenko:2020:square}
M.~Andriushchenko, F.~Croce, N.~Flammarion, and M.~Hein, ``Square attack: {A} query-efficient black-box adversarial attack via random search,'' in \emph{Computer Vision - {ECCV} 2020 - 16th European Conference, Glasgow, UK}, ser. Lecture Notes in Computer Science.\hskip 1em plus 0.5em minus 0.4em\relax Springer, 2020.

\bibitem{croce:2021:robustbench}
F.~Croce, M.~Andriushchenko, V.~Sehwag, E.~Debenedetti, N.~Flammarion, M.~Chiang, P.~Mittal, and M.~Hein, ``Robustbench: a standardized adversarial robustness benchmark,'' in \emph{Proceedings of the Neural Information Processing Systems Track on Datasets and Benchmarks 1, NeurIPS Datasets and Benchmarks 2021}, 2021.

\bibitem{robey:2022:probabilistically}
A.~Robey, L.~F.~O. Chamon, G.~J. Pappas, and H.~Hassani, ``Probabilistically robust learning: Balancing average and worst-case performance,'' in \emph{International Conference on Machine Learning, ({ICML})}, ser. Proceedings of Machine Learning Research.\hskip 1em plus 0.5em minus 0.4em\relax {PMLR}, 2022.

\bibitem{oliver:2023:howmany}
R.~Olivier and B.~Raj, ``How many perturbations break this model? evaluating robustness beyond adversarial accuracy,'' in \emph{International Conference on Machine Learning, ({ICML})}, ser. Proceedings of Machine Learning Research.\hskip 1em plus 0.5em minus 0.4em\relax {PMLR}, 2023.

\bibitem{xu:2018:feature}
W.~Xu, D.~Evans, and Y.~Qi, ``Feature squeezing: Detecting adversarial examples in deep neural networks,'' in \emph{25th Annual Network and Distributed System Security Symposium, {NDSS}}.\hskip 1em plus 0.5em minus 0.4em\relax The Internet Society, 2018.

\bibitem{bader:2011:hype}
J.~Bader and E.~Zitzler, ``Hype: An algorithm for fast hypervolume-based many-objective optimization,'' \emph{Evol. Comput.}, 2011.

\bibitem{deb:2016:multi}
K.~Deb, K.~Sindhya, and J.~Hakanen, ``Multi-objective optimization,'' in \emph{Decision sciences}.\hskip 1em plus 0.5em minus 0.4em\relax CRC Press, 2016, pp. 161--200.

\bibitem{deb:2002:nsga2}
K.~Deb, S.~Agrawal, A.~Pratap, and T.~Meyarivan, ``A fast and elitist multiobjective genetic algorithm: {NSGA-II},'' \emph{{IEEE} Trans. Evol. Comput.}, 2002.

\bibitem{zhang:2007:moead}
Q.~Zhang and H.~Li, ``{MOEA/D:} {A} multiobjective evolutionary algorithm based on decomposition,'' \emph{{IEEE} Trans. Evol. Comput.}, 2007.

\bibitem{akhtar:2018:threat}
N.~Akhtar and A.~S. Mian, ``Threat of adversarial attacks on deep learning in computer vision: {A} survey,'' \emph{{IEEE} Access}, 2018.

\bibitem{machado:2023:adversarial}
G.~R. Machado, E.~Silva, and R.~R. Goldschmidt, ``Adversarial machine learning in image classification: {A} survey toward the defender's perspective,'' \emph{{ACM} Comput. Surv.}, 2023.

\bibitem{li:2023:trustworthy}
B.~Li, P.~Qi, B.~Liu, S.~Di, J.~Liu, J.~Pei, J.~Yi, and B.~Zhou, ``Trustworthy {AI:} from principles to practices,'' \emph{{ACM} Comput. Surv.}, 2023.

\bibitem{bhambri:2019:survey}
S.~Bhambri, S.~Muku, A.~Tulasi, and A.~B. Buduru, ``A survey of black-box adversarial attacks on computer vision models,'' \emph{arXiv preprint arXiv:1912.01667}, 2019.

\bibitem{li:2023:saes}
\BIBentryALTinterwordspacing
Z.~Li, H.~Cheng, X.~Cai, J.~Zhao, and Q.~Zhang, ``{SA-ES:} subspace activation evolution strategy for black-box adversarial attacks,'' \emph{{IEEE} Trans. Emerg. Top. Comput. Intell.}, vol.~7, no.~3, pp. 780--790, 2023. [Online]. Available: \url{https://doi.org/10.1109/TETCI.2022.3214627}
\BIBentrySTDinterwordspacing

\bibitem{li:2020:qeba}
\BIBentryALTinterwordspacing
H.~Li, X.~Xu, X.~Zhang, S.~Yang, and B.~Li, ``{QEBA:} query-efficient boundary-based blackbox attack,'' \emph{CoRR}, vol. abs/2005.14137, 2020. [Online]. Available: \url{https://arxiv.org/abs/2005.14137}
\BIBentrySTDinterwordspacing

\bibitem{fu:2022:autoda}
\BIBentryALTinterwordspacing
Q.~Fu, Y.~Dong, H.~Su, J.~Zhu, and C.~Zhang, ``Autoda: Automated decision-based iterative adversarial attacks,'' in \emph{31st {USENIX} Security Symposium, {USENIX} Security 2022, Boston, MA, USA, August 10-12, 2022}, K.~R.~B. Butler and K.~Thomas, Eds.\hskip 1em plus 0.5em minus 0.4em\relax {USENIX} Association, 2022, pp. 3557--3574. [Online]. Available: \url{https://www.usenix.org/conference/usenixsecurity22/presentation/fu-qi}
\BIBentrySTDinterwordspacing

\bibitem{guo:2024:lautoda}
\BIBentryALTinterwordspacing
P.~Guo, F.~Liu, X.~Lin, Q.~Zhao, and Q.~Zhang, ``L-autoda: Leveraging large language models for automated decision-based adversarial attacks,'' \emph{CoRR}, vol. abs/2401.15335, 2024. [Online]. Available: \url{https://doi.org/10.48550/arXiv.2401.15335}
\BIBentrySTDinterwordspacing

\bibitem{zanddizari:2022:gener}
\BIBentryALTinterwordspacing
H.~Zanddizari, B.~Zeinali, and J.~M. Chang, ``Generating black-box adversarial examples in sparse domain,'' \emph{{IEEE} Trans. Emerg. Top. Comput. Intell.}, vol.~6, no.~4, pp. 795--804, 2022. [Online]. Available: \url{https://doi.org/10.1109/TETCI.2021.3122467}
\BIBentrySTDinterwordspacing

\bibitem{mao:2021:composite}
X.~Mao, Y.~Chen, S.~Wang, H.~Su, Y.~He, and H.~Xue, ``Composite adversarial attacks,'' in \emph{Thirty-Fifth {AAAI} Conference on Artificial Intelligence, {AAAI}}, 2021.

\bibitem{liu:2023:reliable}
S.~Liu, F.~Peng, and K.~Tang, ``Reliable robustness evaluation via automatically constructed attack ensembles,'' in \emph{Thirty-Seventh {AAAI} Conference on Artificial Intelligence, {AAAI}}.\hskip 1em plus 0.5em minus 0.4em\relax {AAAI} Press, 2023.

\bibitem{bai:2021:recent}
T.~Bai, J.~Luo, J.~Zhao, B.~Wen, and Q.~Wang, ``Recent advances in adversarial training for adversarial robustness,'' in \emph{Proceedings of the Thirtieth International Joint Conference on Artificial Intelligence, ({IJCAI})}.\hskip 1em plus 0.5em minus 0.4em\relax ijcai.org, 2021.

\bibitem{ye:2024:shed}
Y.~He, Z.~Wang, Z.~Shen, G.~Sun, Y.~Dai, Y.~Wu, H.~Wang, and A.~Li, ``{SHED:} shapley-based automated dataset refinement for instruction fine-tuning,'' in \emph{Advances in Neural Information Processing Systems 37: Annual Conference on Neural Information Processing Systems 2024, NeurIPS}, 2024.

\bibitem{wang:2024:flora}
Z.~Wang, Z.~Shen, Y.~He, G.~Sun, H.~Wang, L.~Lyu, and A.~Li, ``{FLoRA}: Federated fine-tuning large language models with heterogeneous low-rank adaptations,'' in \emph{Advances in Neural Information Processing Systems 37: Annual Conference on Neural Information Processing Systems 2024, NeurIPS}, 2024.

\bibitem{liu:2023:twins}
Z.~Liu, Y.~Xu, X.~Ji, and A.~B. Chan, ``{TWINS:} {A} fine-tuning framework for improved transferability of adversarial robustness and generalization,'' in \emph{{IEEE/CVF} Conference on Computer Vision and Pattern Recognition, {CVPR}}.\hskip 1em plus 0.5em minus 0.4em\relax {IEEE}, 2023.

\bibitem{guo:2018:countering}
C.~Guo, M.~Rana, M.~Ciss{\'{e}}, and L.~van~der Maaten, ``Countering adversarial images using input transformations,'' in \emph{6th International Conference on Learning Representations, ({ICLR})}.\hskip 1em plus 0.5em minus 0.4em\relax OpenReview.net, 2018.

\bibitem{guo:2024:puridefense}
P.~Guo, Z.~Yang, X.~Lin, Q.~Zhao, and Q.~Zhang, ``Puridefense: Randomized local implicit adversarial purification for defending black-box query-based attacks,'' \emph{CoRR}, 2024.

\bibitem{carlini:2023:certified}
N.~Carlini, F.~Tram{\`{e}}r, K.~D. Dvijotham, L.~Rice, M.~Sun, and J.~Z. Kolter, ``(certified!!) adversarial robustness for free!'' in \emph{The Eleventh International Conference on Learning Representations, ({ICLR})}.\hskip 1em plus 0.5em minus 0.4em\relax OpenReview.net, 2023.

\bibitem{ma:2019:nic}
S.~Ma, Y.~Liu, G.~Tao, W.~Lee, and X.~Zhang, ``{NIC:} detecting adversarial samples with neural network invariant checking,'' in \emph{26th Annual Network and Distributed System Security Symposium, ({NDSS})}.\hskip 1em plus 0.5em minus 0.4em\relax The Internet Society, 2019.

\bibitem{li:2022:blacklight}
H.~Li, S.~Shan, E.~Wenger, J.~Zhang, H.~Zheng, and B.~Y. Zhao, ``Blacklight: Scalable defense for neural networks against query-based black-box attacks,'' in \emph{31st {USENIX} Security Symposium, {USENIX} Security 2022}.\hskip 1em plus 0.5em minus 0.4em\relax {USENIX} Association, 2022.

\bibitem{zhang:2019:theoretically}
H.~Zhang, Y.~Yu, J.~Jiao, E.~P. Xing, L.~E. Ghaoui, and M.~I. Jordan, ``Theoretically principled trade-off between robustness and accuracy,'' in \emph{Proceedings of the 36th International Conference on Machine Learning, ({ICML})}, ser. Proceedings of Machine Learning Research.\hskip 1em plus 0.5em minus 0.4em\relax {PMLR}, 2019.

\bibitem{wang:2023:better}
Z.~Wang, T.~Pang, C.~Du, M.~Lin, W.~Liu, and S.~Yan, ``Better diffusion models further improve adversarial training,'' in \emph{International Conference on Machine Learning, ({ICML})}, ser. Proceedings of Machine Learning Research.\hskip 1em plus 0.5em minus 0.4em\relax {PMLR}, 2023.

\bibitem{gowal:2020:uncovering}
S.~Gowal, C.~Qin, J.~Uesato, T.~A. Mann, and P.~Kohli, ``Uncovering the limits of adversarial training against norm-bounded adversarial examples,'' \emph{CoRR}, 2020.

\bibitem{williams:2023:blackbox}
P.~N. Williams and K.~Li, ``Black-box sparse adversarial attack via multi-objective optimisation {CVPR} proceedings,'' in \emph{{IEEE/CVF} Conference on Computer Vision and Pattern Recognition, {CVPR}}.\hskip 1em plus 0.5em minus 0.4em\relax {IEEE}, 2023.

\bibitem{williams:2023:sparse}
P.~N. Williams, K.~Li, and G.~Min, ``Sparse adversarial attack via bi-objective optimization,'' in \emph{Evolutionary Multi-Criterion Optimization - 12th International Conference, {EMO}}.\hskip 1em plus 0.5em minus 0.4em\relax Springer, 2023.

\bibitem{deist:2023:multi}
T.~M. Deist, M.~Grewal, F.~J. W.~M. Dankers, T.~Alderliesten, and P.~A.~N. Bosman, ``Multi-objective learning using {HV} maximization,'' in \emph{Evolutionary Multi-Criterion Optimization - 12th International Conference, {EMO}}.\hskip 1em plus 0.5em minus 0.4em\relax Springer, 2023.

\bibitem{Suzuki:2019:adversarial}
T.~Suzuki, S.~Takeshita, and S.~Ono, ``Adversarial example generation using evolutionary multi-objective optimization,'' in \emph{{IEEE} Congress on Evolutionary Computation, {CEC}}.\hskip 1em plus 0.5em minus 0.4em\relax {IEEE}, 2019.

\bibitem{baia:2021:effective}
A.~E. Baia, G.~D. Bari, and V.~Poggioni, ``Effective universal unrestricted adversarial attacks using a {MOE} approach,'' in \emph{Applications of Evolutionary Computation - 24th International Conference, EvoApplications}, ser. Lecture Notes in Computer Science.\hskip 1em plus 0.5em minus 0.4em\relax Springer, 2021.

\bibitem{liu:2024:effective}
S.~Liu, N.~Lu, W.~Hong, C.~Qian, and K.~Tang, ``Effective and imperceptible adversarial textual attack via multi-objectivization,'' \emph{ACM Trans. Evol. Learn. Optim.}, 2024.

\bibitem{CIFAR10}
A.~Krizhevsky, ``{Learning Multiple Layers of Features from Tiny Images},'' Univ. Toronto, Technical Report, 2009.

\bibitem{rade:2021:helperbased}
\BIBentryALTinterwordspacing
R.~Rade and S.-M. Moosavi-Dezfooli, ``Helper-based adversarial training: Reducing excessive margin to achieve a better accuracy vs. robustness trade-off,'' in \emph{ICML 2021 Workshop on Adversarial Machine Learning}, 2021. [Online]. Available: \url{https://openreview.net/forum?id=BuD2LmNaU3a}
\BIBentrySTDinterwordspacing

\bibitem{sehwag:2022:robust}
V.~Sehwag, S.~Mahloujifar, T.~Handina, S.~Dai, C.~Xiang, M.~Chiang, and P.~Mittal, ``Robust learning meets generative models: Can proxy distributions improve adversarial robustness?'' in \emph{The Tenth International Conference on Learning Representations, ({ICLR})}.\hskip 1em plus 0.5em minus 0.4em\relax OpenReview.net, 2022.

\bibitem{augustin:2020:adversarial}
M.~Augustin, A.~Meinke, and M.~Hein, ``Adversarial robustness on in- and out-distribution improves explainability,'' in \emph{Computer Vision - {ECCV} 2020 - 16th European Conference}, ser. Lecture Notes in Computer Science.\hskip 1em plus 0.5em minus 0.4em\relax Springer, 2020.

\bibitem{nowak:2014:empirical}
K.~Nowak, M.~M{\"{a}}rtens, and D.~Izzo, ``Empirical performance of the approximation of the least hypervolume contributor,'' in \emph{Parallel Problem Solving from Nature - {PPSN} {XIII} - 13th International Conference, Ljubljana, Slovenia, September 13-17, 2014. Proceedings}, ser. Lecture Notes in Computer Science.\hskip 1em plus 0.5em minus 0.4em\relax Springer, 2014, pp. 662--671.

\bibitem{while:2012:fast}
L.~While, L.~Bradstreet, and L.~Barone, ``A fast way of calculating exact hypervolumes,'' \emph{{IEEE} Trans. Evol. Comput.}, 2012.

\bibitem{addepalli:2022:efficient}
S.~Addepalli, S.~Jain, and V.~B. R., ``Efficient and effective augmentation strategy for adversarial training,'' in \emph{Advances in Neural Information Processing Systems 35: Annual Conference on Neural Information Processing Systems 2022, ({NeurIPS})}, 2022.

\bibitem{rade:2022:reducing}
R.~Rade and S.~Moosavi{-}Dezfooli, ``Reducing excessive margin to achieve a better accuracy vs. robustness trade-off,'' in \emph{The Tenth International Conference on Learning Representations, ({ICLR})}.\hskip 1em plus 0.5em minus 0.4em\relax OpenReview.net, 2022.

\bibitem{rebuffi:2021:fixing}
S.~Rebuffi, S.~Gowal, D.~A. Calian, F.~Stimberg, O.~Wiles, and T.~A. Mann, ``Fixing data augmentation to improve adversarial robustness,'' \emph{CoRR}, 2021.

\bibitem{huang:2021:exploring}
H.~Huang, Y.~Wang, S.~M. Erfani, Q.~Gu, J.~Bailey, and X.~Ma, ``Exploring architectural ingredients of adversarially robust deep neural networks,'' in \emph{Advances in Neural Information Processing Systems 34: Annual Conference on Neural Information Processing Systems 2021, ({NeurIPS})}, 2021.

\bibitem{imagenet_cvpr09}
J.~Deng, W.~Dong, R.~Socher, L.-J. Li, K.~Li, and L.~Fei-Fei, ``{ImageNet: A Large-Scale Hierarchical Image Database},'' in \emph{CVPR09}, 2009.

\bibitem{wong:2020:fast}
E.~Wong, L.~Rice, and J.~Z. Kolter, ``Fast is better than free: Revisiting adversarial training,'' in \emph{8th International Conference on Learning Representations, {ICLR}}.\hskip 1em plus 0.5em minus 0.4em\relax OpenReview.net, 2020.

\bibitem{salman:2020:do}
H.~Salman, A.~Ilyas, L.~Engstrom, A.~Kapoor, and A.~Madry, ``Do adversarially robust imagenet models transfer better?'' in \emph{Advances in Neural Information Processing Systems 33: Annual Conference on Neural Information Processing Systems 2020, NeurIPS}, 2020.

\bibitem{robustness}
\BIBentryALTinterwordspacing
L.~Engstrom, A.~Ilyas, H.~Salman, S.~Santurkar, and D.~Tsipras, ``Robustness (python library),'' 2019. [Online]. Available: \url{https://github.com/MadryLab/robustness}
\BIBentrySTDinterwordspacing

\bibitem{diamond:2016:cvxpy}
S.~Diamond and S.~Boyd, ``{CVXPY}: {A} {P}ython-embedded modeling language for convex optimization,'' \emph{Journal of Machine Learning Research}, vol.~17, no.~83, pp. 1--5, 2016.

\bibitem{brown:2017:adversarial}
T.~B. Brown, D.~Man{\'{e}}, A.~Roy, M.~Abadi, and J.~Gilmer, ``Adversarial patch,'' in \emph{Advances in Neural Information Processing Systems 31: Annual Conference on Neural Information Processing Systems 2017, NeurIPS}, 2017.

\end{thebibliography}

\vfill

\end{document}